\documentclass[11pt,twoside]{article}
\usepackage{asp2010}
\resetcounters

\begin{document}

\title{The Massive Star Content of Circumnuclear Star Clusters in M83}
\author{Aida Wofford$^1$, Rupali Chandar$^2$, and Claus Leitherer$^1$
\affil{$^1$Space Telescope Science Institute, 3700 San Martin Drive, Baltimore, MD 21218, USA}
\affil{$^2$University of Toledo, Department of Physics and Astronomy, Toledo, OH 43606, USA}}

\begin{abstract}
The circumnuclear starburst of M83 (NGC 5236), the nearest such example (4.6 Mpc), constitutes an ideal site for studying the massive star IMF at high metallicity (12+log[O/H]=9.1$\pm$0.2, \citealt{bre02}). We analyzed archival \textit{HST}/STIS FUV imaging and spectroscopy of 13 circumnuclear star clusters in M83. We compared the observed spectra with two types of single stellar population (SSP) models, semi-empirical models, which are based on an empirical library of Galactic O and B stars observed with \textit{IUE} \citep{rob93}, and theoretical models, which are based on a new theoretical UV library of hot massive stars described in \cite{lei10} and computed with WM-Basic \citep{pau01}. The models were generated with Starburst99 \citep{lei09}. We derived the reddenings, the ages, and the masses of the clusters from model fits to the FUV spectroscopy, as well as from optical \textit{HST}/WFC3 photometry.
\end{abstract}

\section{Observations}
M83 is a southern, nearly face-on, grand-design spiral galaxy, hosting an arc-shaped circumnuclear starburst of $\sim$200\,pc in length and $\sim$35\,pc in thickness. The starburst is composed of several dozen clusters, which are located within a 200\,pc of M83's optical nucleus. We analyzed 1200-1700\,\AA~52''$\times$2" slit spectroscopy in for nine bright compact clusters, as well as 1200-1700\,\AA~slitless spectrograms, for four additional clusters,  all 
within M83's circumnuclear starburst. The FUV data and their analysis are described in \cite{wof10}. We compared the cluster ages and masses derived from the FUV spectroscopy with values obtained from optical \textit{HST}/WFC3 photometry and derived as described in \cite{cha10}.

\section{Model Spectra}
\begin{figure}[!ht]
\plotone{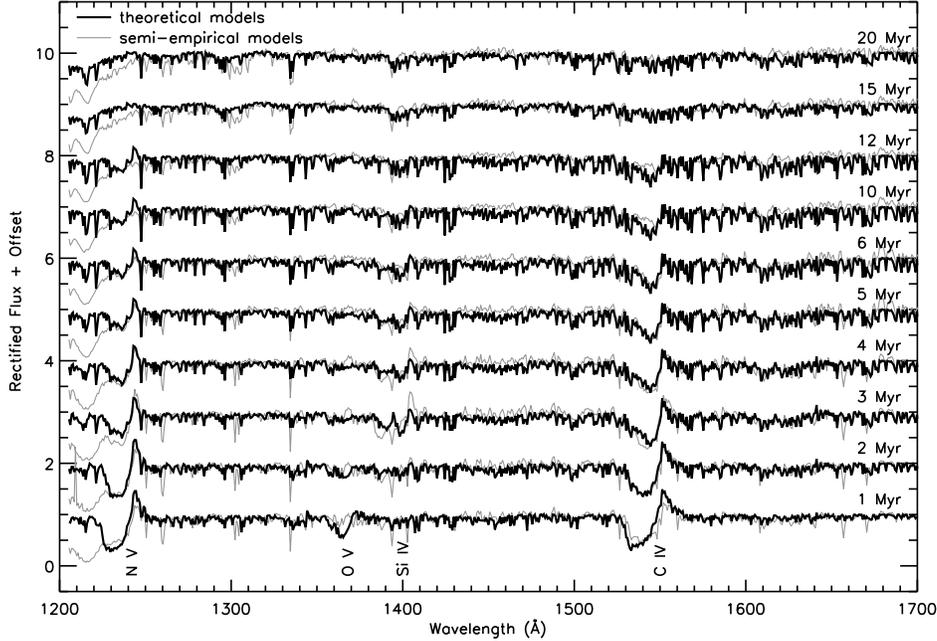}
\caption{Evolution of SSP model spectra with time. The theoretical models are shown in black and the semi-empirical models are shown in grey. The models correspond to a metallicity of $Z=0.020$ and a Kroupa IMF from 0.1-100 M$_\odot$.
\label{fig:models}
}
\end{figure}

\cite{bre02} ruled out the possibility that star formation within individual clusters proceeded continuously in the circumnuclear starburst of M83. Therefore, we fitted our cluster spectra with SSP models. Figure~\ref{fig:models} shows the age evolution of SSP models from 1 to 20\,Myr for models based on the empirical and the theoretical stellar libraries, hereafter referred to as the semi-empirical and the theoretical models, respectively. The semi-empirical and theoretical  models differ below 1240\,\AA~because of the presence of Galactic interstellar Ly$\alpha$ but agree rather well in the range 1240-1700\,\AA, except for the O V 1370\,\AA~line, which is stronger in the theoretical models at ages younger than $\sim$2\,Myr, and the Si IV 1400\,\AA~feature, which is stronger in the semi-empirical models at ages 3 and 4\,Myr. Note that the empirical library is contaminated with interstellar lines.

\section{Procedure}
\subsection{Reddening}
We corrected the observed spectra for Galactic extinction based on the maps of \cite{sch98} and the extinction curve of \cite{fit99}. We then fitted the FUV continuum with a power law of the form F$\sim$$\lambda^\beta$ and assumed that any deviation of $\beta$ from the expected value for a dust free starburst (-2.6) was due to reddening. We used the obscuration law of \cite{cal00} to deredden the spectrum until $\beta$=-2.6 was reached.
\subsection{Spectroscopic Ages, Metallicity, and IMF}
\begin{figure}[!ht]
\plotone{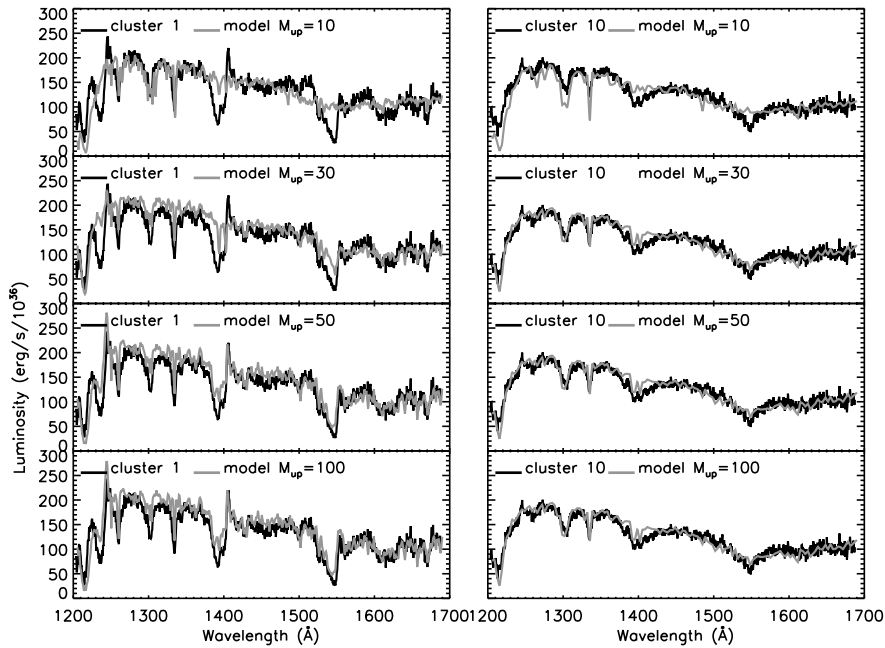}
\caption{Observed spectra of clusters 1 and 10 (black curves) versus semi-empirical models corresponding to different upper mass limits to the IMF, M$_\textrm{up}=10$, 30, 50, and 100\,M$_\odot$ (grey curves). The metallicity of the models is $Z=0.020$.  The ages of clusters 1 and 10 are $\sim$4\,Myr and $\sim$12\,Myr, while their masses are $\sim$3$\times 10^4$\,M$_\odot$ and  $\sim$$10^5$\,M$_\odot$, respectively. 
\label{fig:upimf}
}
\end{figure}

Figure~\ref{fig:models} shows the sensitivity of the N V 1240, Si IV 1400, and C IV 1550 profiles to the cluster age. M83's starburst metallicity is intermediate between $Z=0.020$ and $Z=0.040$, but the age estimates from models with these two metallicities are very similar. Our clusters showing strong P-Cygni profiles in N V 1240, Si IV 1400, and C IV 1550 are consistent with having formed stars more massive than 30\,M$_\odot$. The presence of Si IV and C IV absorption in the rest of clusters (assuming that it is not interstellar), suggests the presence of at least some B stars. This is illustrated in Fig.~\ref{fig:upimf}, where the observed spectra of clusters 1 and 10 are compared against models having upper mass limits of 10, 30, 50, and 100\,M$_\odot$. We derived spectroscopic ages for all the clusters in our sample by fitting their FUV spectra with models having a metallicity of $Z=0.020$ and a Kroupa IMF from 0.1-100\,M$_\odot$. We found the best fit model by giving the most weight to N V 1240, Si IV 1400, and C IV 1550.
\subsection{Spectroscopic Masses}
The spectroscopic mass of each cluster was derived from the mean luminosity of the observed FUV continuum. 

\section{CLUSTER PROPERTIES}
\subsection{Masses}
Optical photometry provides more leverage for determining the stellar mass than FUV spectroscopy. Therefore, our photometric masses (M$_p$) are more reliable than our spectroscopic masses. Our most massive cluster has M$_p=3.1\times10^5$\,M$_\odot$, which is comparable to the virial mass of the ionizing cluster of 30 Doradus, NGC 2070 ($4.5\times10^5$\,M$_\odot$, \citealt{bos09}). According to \cite{lar10}, young star clusters with masses larger than 10$^5$\,M$_\odot$ can last an age comparable or exceeding the age of the universe. Therefore the latter cluster could be a globular cluster progenitor, while two other clusters, which have M$_p\approx10^5$\,M$_\odot$, are just in the limit. One cluster has M$_p=4\times10^3$\,M$_\odot$. Unfortunately, its spectrum is too noisy to reliably say whether stars more massive than 30\,M$_\odot$ have formed in it. The rest of clusters have M$_p$ of a few$\,\times10^4$\,M$_\odot$.
\subsection{IMF}
Our clusters with strong P-Cygni profiles have M$_p\sim10^4$\,M$_\odot$, seem to have formed stars with masses $>30$\,M$_\odot$, and are consistent with a Kroupa IMF from 0.1-100 M$_\odot$. The rest of clusters are consistent with having formed at least some B stars.
\subsection{Ages}
The spectroscopic ages from semi-empirical and theoretical predictions are within a factor of 1.2.  The spectroscopic and photometric ages agree at a similar level. Our ages agree with those derived from \textit{HST}/WFPC2 photometry by \cite{har01}, except for clusters 6, 7, and 10, which are older than 6\,Myr in our case. Our ages for clusters 1-3 agree with the ages of region A derived from STIS FUV spectroscopy by \cite{bre02}. The clusters are $\sim$3-20\,Myr old and were not all formed at the same time. We found no age gradient along M83's starburst, in disagreement with \cite{pux97} and \cite{dia06}.

\acknowledgements This work was supported by NASA grant N1317.

\bibliography{wofford_a_ref}

\end{document}